\documentclass[conference,letterpaper]{IEEEtran}
\IEEEoverridecommandlockouts
\usepackage{cite}
\usepackage{amsmath,amssymb,amsfonts}
\usepackage{algorithmic}
\usepackage{graphicx}
\usepackage{textcomp}
\usepackage{xcolor}
\usepackage{subcaption}
\usepackage{soul}
\usepackage[normalem]{ulem}
\usepackage{url}
\sethlcolor{yellow}
\usepackage{fancyhdr}

\newif\ifcomments
\commentsfalse

\definecolor{darkgreen}{RGB}{0,95,0} % classic dark green
\definecolor{ultramarine}{rgb}{0.07, 0.04, 0.76}
\definecolor{brickred}{rgb}{0.8, 0.25, 0.33}

\renewcommand{\hl}[1]{#1} % added to remove highlighting

\begin{document}
\bstctlcite{IEEEexample:BSTcontrol}
\title{Advanced LLM-Enhanced Intent-Based 5G Network Management using Dynamic Semantic Routes}

\author{Thomas~Benton~Townsend and~Dimitrios~Michael~Manias \\ The Department of Computer Science and Engineering, Mississippi State University\\ tbt135@msstate.edu, dmanias@cse.msstate.edu}% <-this % stops a space

\maketitle

\begin{abstract}
As the use of Artificial Intelligence (AI) and Large Language Models (LLMs) is becoming common in everyday applications, their ability to interpret natural language has increased significantly. An emerging application of AI is integration with network management and orchestration practices. An instance of this integration  is LLM-enhanced intent-based networking, where network operators will control a network using natural language. This work presents the use of dynamic routes with a semantic router to identify an intent from a network operator's prompt and extract necessary details for intent fulfillment in intent-based 5G+ core networks. Furthermore, the performance of static route selection is assessed by evaluating  multiple encoders and dynamic route detail extraction accuracy against a series of realistic  operator prompts. The presented results show that static and dynamic routes are successful in detail extraction and schema formatting.  

\end{abstract}


\section{Introduction}
As modern networks continue to evolve through the introduction of new generations (\textit{e.g.,} 5G+), the development of modern network configurations (\textit{e.g.,} IoT), and the realization of paradigm-shifting technologies (\textit{e.g.,} multi-access edge computing, network slicing), significant challenges emerge in their management\cite{10230112}. Modern networks are growing at an unprecedented pace, both in terms of the number of users and devices, and in terms of expected performance and requirements. Given that networks are becoming increasingly distributed and dynamic, human-centric network management is becoming increasingly infeasible due to scale and complexity. Operators are tasked with determining a suitable replacement for human-centered management that will endure across a constantly advancing, unpredictable network landscape.

It is a widely accepted view in the field that Zero-touch network and Service Management (ZSM) is the end goal. ZSM describes a vision of network automation in which networks exhibit key properties, including self-configuration, self-optimization, and self-healing \cite{ZSM}. While the envisioned architecture is \textit{Human-out-of-the-Loop}, there are several stages that current \textit{Human-in-the-Loop} network management practices must advance through before this vision of full automation can be realized. A natural stage for this transition is \textit{Human-on-the-Loop} management, where the network operator conveys goals and requirements (intents), and the network itself can act on them. This type of networking is referred to as Intent-based Networking.

One of the requirements of \textit{Human-on-the-Loop} networking is the translation of natural language intents into machine-readable actions. The proliferation of accessible Machine Learning (ML) and Artificial Intelligence (AI) models, specifically Large Language Models (LLMs), capable of processing natural language inputs, presents a unique opportunity for integration with Intent-based Networking. This integration helps address the increasingly complex nature of modern network management by moving from a \textit{Human-in-the-Loop} to a \textit{Human-on-the-Loop} approach, with the ultimate goal of full automation through ZSM becoming increasingly attainable.

To this end, the work presented in this paper builds upon past work and proposes an enhanced LLM-based Intent-based networking approach for 5G+ Core (5GC) network management. Previously, a set of intents with expectations for the 5GC was defined, and initial intent identification was conducted \cite{10539172}. Subsequent work addressed the hallucination phenomenon that commonly manifests in LLM implementations by introducing static semantic routes, thereby making the system behave deterministically \cite{10901065}. This work advances prior work by improving the quality of intent identification and transitioning from static to dynamic routing. The use of dynamic routing introduces novel capabilities for extracting and formatting key parameters from the user intent into a policy schema for direct use in the network.  

The rest of this paper is organized as follows: Section II reviews the state of the art; Section III introduces the system model; Section IV describes the methodology; Section V analyzes the results; and Section VI concludes the paper with a discussion on future work.

\section{Related Work}

LLMs are praised for their ability to interpret, translate, and generate a variety of modalities, including natural language, structured code, and high-dimensional time-series telemetry data. Naturally, the complexity of modern networks and systems presents a unique opportunity to integrate LLMs for enhanced management, orchestration, and automation. 

Liu \textit{et al.} \cite{liu2025large} synthesize network-related LLM integrations into four broad categories: design, optimization, security, and automation. 
In the context of network design, LLMs have been used for tasks ranging from generating router configurations \cite{mondal2023llms} to designing network algorithms \cite{he2024designing}. Regarding optimization, LLMs have been leveraged to enhance performance in various network architectures, including optical networks \cite{song2025synergistic} and core networks \cite{li2025next}. In the context of security, LLMs are increasingly being leveraged for anomaly-detection tasks, most commonly for network intrusion detection \cite{11059643}. Despite the increasing popularity of LLMs in these domains, the field of network automation, specifically intent-based networking, has seen the most rapid proliferation of LLM integration. 

Intent-based network management has gained significant traction in recent years due to the advancements in natural language processing and the proliferation of open, accessible LLMs. Leivadeas and Falkner \cite{9925251} present a comprehensive survey on intent-based networking, focusing on a closed-loop automation cycle comprising intent profiling, translation, resolution, activation, and assurance. The authors propose methods, including templates, graphical user interfaces, natural language processing, specific intent-based languages, and machine learning, to handle the various stages of the intent lifecycle. Mekrache \textit{et al.} \cite{10574890} propose the use of open-source LLMs to achieve end-to-end intent lifecycle management. The authors develop a framework that decomposes natural language into domain-specific intent components and translates it into structured outputs such as JSON. The authors determined that the Code Llama 34B outperformed other models; however, it still faced processing challenges as intent volume and complexity increased.

In previous work, Manias \textit{et al.} established six intent types that include an expectation for the 5GC network, based on the 3GPP Technical Specification 28.312 (ETSI TS 128 312) \cite{10539172}. The authors focused on intent extraction and identified two key limitations with network-related LLM integration: (1) the stochastic output variability of LLMs leading to hallucinations, and (2) a reliance on proprietary, closed-source models. To address these limitations, subsequent work used a semantic router to extract intent from user input and trigger a static route, along with open-source LLMs, namely Mistral 7B \cite{10901065}. While this enabled deterministic behavior, it lacked the flexibility required for complex, evolving network states.

The work presented in this paper builds on prior work by converting the semantic router’s static routes into dynamic ones. The presented system translates natural-language intents into machine-executable policies, moving another step toward zero-touch network management and real-time provisioning and enforcement across an operational network. 

The contributions of this work are summarized as follows:
\begin{itemize}
\item An end-to-end intent extraction model that uses a semantic router with associated dynamic routes to realize intent-based networking in the 5GC.
\item The creation of static routes and dynamic route functions that generate a machine-readable schema output.
\item A comparative analysis of five encoder options for intent extraction and route selection.
\end{itemize}

\section{System Model}

\begin{figure*}[!hbt]
\centerline{\includegraphics[width=1.75\columnwidth]{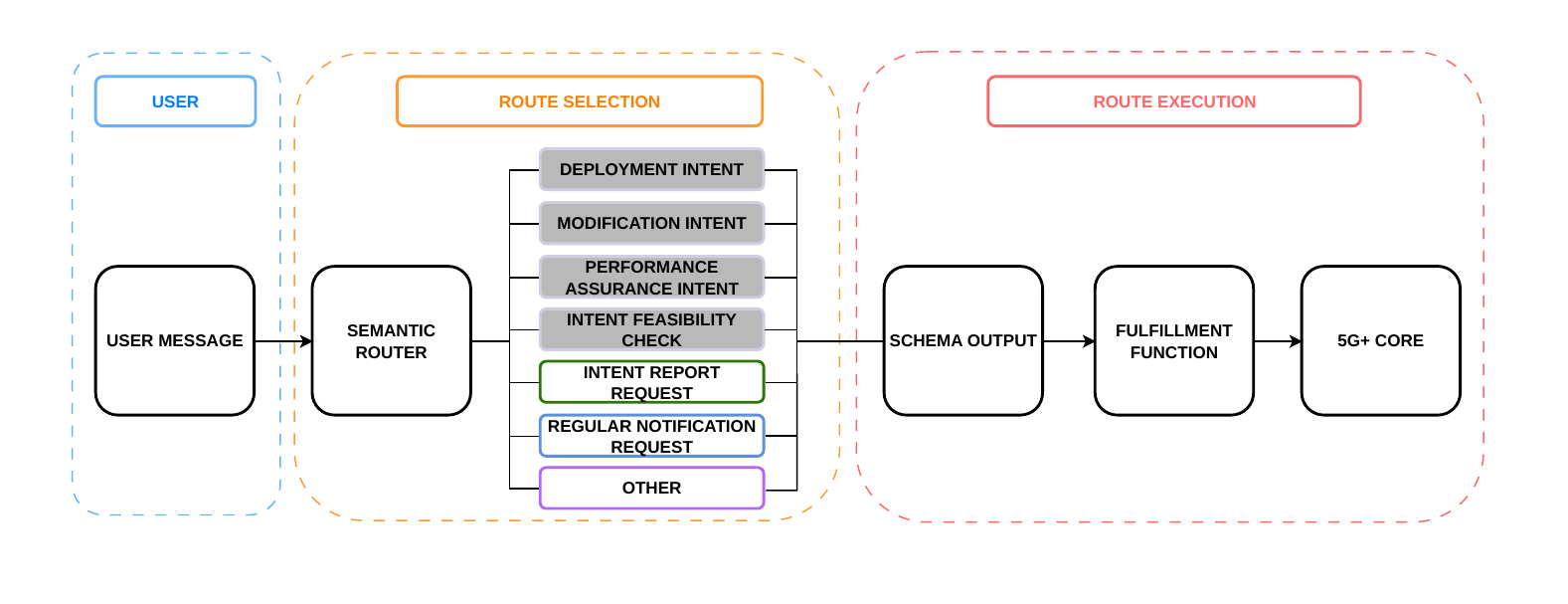}}
\caption{High-Level System Overview with Subsystem Labeling}
\label{fig:system_model}
\end{figure*}

The proposed system model depicted in Fig. \ref{fig:system_model} has three subsystems: the user, the route selection, and the route execution. The \hl{interaction with the system is initiated by the} user’s message. Firstly, the user will send a message to the semantic router. \hl{Because the user will be a network operator with system knowledge, }it is assumed that the message will be formatted in a specific way depending on the operator's intent. This allows for the proper extraction of relevant values in the route execution. Second, the semantic router will select a route based on the intent extracted from the user’s message. This system focuses on the Regular Notification Request (RNR) and Intent Report Request (IRR) intents. Lastly, the selected route will execute a dynamic route. The dynamic route uses a predefined function to parse relevant details from the user’s message. These details are then formatted into a JSON-like structure for preparation of user intent fulfillment. The objective of this work is to effectively extract both the user’s intent and generate a schema output with the relevant details for intent fulfillment. Because of this, this work focuses on the RNR and IRR as their relevant details for fulfillment are \hl{concise}. These intents
demonstrate the functionality of the proposed approach and
serve as a foundation to consider more complex intent and schema structures
in future work.
% The small structure of the output was specifically chosen to promote the creation of baseline dynamic routes, thus creating the foundation for future endeavors in this topic. 

\textbf{User Input:}
For each intent used in this work, it is assumed that the user’s input will contain key details that are needed to create a schema output. These assumptions enable proper definition of both RNR and IRR utterances, training, and testing data. If the user wants to be notified regularly about the status of a certain network function, this work assumes that the user will always provide a subject and frequency for notification. Similarly, if the user wants to know the status of a previous intent, it is assumed that the user will provide an intent identifier or a phrase that is parsed as a positional statement. This would include a phrase such as “latest intent” or “last 5 intents”. For this positional statement, it is also assumed the user will only refer to continuous intents starting from the most recent intent. Without these assumptions, it is not a realistic expectation to parse the user’s input nor create a proper schema as an output. \hl{An exception is raised if the system is not able to extract the required values.}

\textbf{Route Selection:}
The semantic router uses a predefined route layer to ingest the user’s message and choose a route. Within the route layer is an encoder that parses the user’s message and generates a similarity score. The thresholds of the route layer and the similarity score is what determines which route is selected. Naturally, different encoders have different performance, and the choice of encoder can have an effect on the route selection. This work focuses on four different encoders: Hugging Face (HF), Alibaba’s Qwen (AB), Mistral (MS), and OpenAI (OA). \hl{In addition to these four standalone encoders, a Hybrid Encoder (HY), which combines a dense and sparse encoder, was also tested.} Their performance before and after training with a variety of different prompts are tested and analyzed. Although the user’s message has an assumed format \hl{(\textit{i.e.,} content),} the composition of the message will differ. The effectiveness of an encoder and its route selection can be tested using these differing prompts. If an encoder can regularly identify a message’s intent and trigger its corresponding route, this encoder is considered to be effective. 

\begin{figure}[!hbt]
\centerline{\includegraphics[width=0.99\columnwidth]{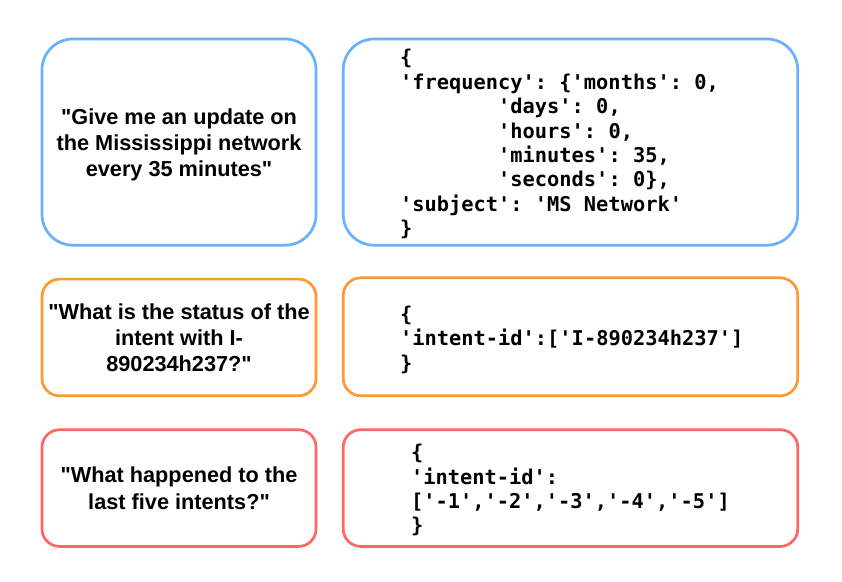}}
\caption{Examples of User Inputs and Resulting Outputs}
\label{fig:rnr_irr_examples}
\end{figure}

\textbf{Route Execution:}
For each intent, a dynamic route is used to execute a function that retrieves key information from the user’s message. As an example, if an RNR route is chosen by the route layer, the resulting schema will contain the network element (\textit{i.e.}, node, link,, function, \textit{etc.}), a partition of the network (\textit{i.e.}, subnet, slice) or the entire network itself to be reminded of and the frequency of notification. The result of an RNR request is seen in Fig. \ref{fig:rnr_irr_examples}. The objective of the IRR dynamic route is to identify the intent identification number or a positional statement that refers to a previous intent or intents. \hl{This work assumes that the user will either provide a valid UUID-based intent identifier or a statement that only refers to intents in a continuous format.} Figure \ref{fig:rnr_irr_examples} shows both intent identifiers and their associated user prompts.

\section{Methodology}
 
% \hl{This section outlines the various methods used in this work.}
\textbf{Prompt Generation:} A base set of user messages were human-written and then augmented using \hl{Gemini and ChatGPT} to alter the length, phrasing, punctuation, and spelling of each prompt. These prompts were created to trigger one of three routes: RNR, IRR, or the default route None. Any message that triggers a None route is one that does not fit the criteria of either the RNR or IRR routes. These None route messages are important as they test the encoder on its ability to properly extract contextual information from the user’s input and differentiate genuine messages from messages with no relevant intent. All generated prompts were used to define utterances (examples) for both the IRR and RNR routes, train each encoder to test performance improvements, and test the overall performance of each encoder. Without proper message variability, the encoder could determine when to trigger a route based on details like punctuation or capitalization rather than truly understanding the semantics behind a user’s intent. Thus, creating this comprehensive set of inputs \hl{is} vital for proper training and testing of each encoder. 

\textbf{Semantic Router Implementation:} Semantic Router is an open-source tool created by Aurelio AI \cite{semantic_router}. In order to use the router, a set of static routes must first be defined. Each of these routes have utterances that define what type of prompts should trigger that route. For each route, a set of forty utterances have been created. \hl{A route will be triggered if the encoder produces a similarity score for the input message that exceeds the route-specific threshold for that route layer.} After the routes are defined, the next step is to define a route layer. Each route layer uses the routes that have already been defined and takes the chosen encoder as an argument. For the purpose of this work, multiple route layers are created with different encoders. After a layer is created, the layer can be called and the user’s prompt is passed as an argument. The resulting output will be the route that the encoder has chosen.

\textbf{Dynamic Route Implementation:}
The dynamic route feature of the Semantic Router facilitates the execution of a pre-defined function using the user’s message as an argument. The dynamic route is automatically performed when its associated static route is chosen. \hl{The function is used to extract key information from the input and generate a key-value pair(s) as a result.} The natural progression of this system will take the resulting schema from the dynamic route and use it to fulfill the user’s requested action. However, this implementation remains out of scope for this work.

\textit{1) Dynamic Route Creation}
To define a dynamic route, an LLM must first be chosen to fulfill the functions that will be defined. This component of the work uses the Mistral 7B model from Ollama’s LLM collection. The goal of the LLM is to follow the dynamic route function’s guidelines and generate the desired result with the user’s message. In order to define a function for the dynamic route, it is given a name and a function schema that is used as a reference for what the function will do. In this model, the function need only contain this schema. No other details are needed in the function for the LLM to generate a proper output. It is worth noting that the effectiveness of the LLMs output weighs heavily on how the function schema is defined and how effectively the model interprets instructions.

% Comparing the effectiveness of different LLMs and how they interpret function schema is not within the scope of this work. \hl{However, the importance of schema interpretation in dynamic route functions justifies this testing in future works.}

% \begin{table}[]
% \begin{tabular}{|c|c|}
% \hline
% \multicolumn{1}{|l|}{\textbf{Regular Notification Request}} & \multicolumn{1}{l|}{\textbf{Intent %Report Request}} \\ \hline
% "Give me the status of AMF every 5 hours." & "What happened with the last 3 intents?" \\ \hline
% "Update about NWDAF every hour" & "Intent I-7723498 update." \\ \hline
% \begin{tabular}[c]{@{}c@{}}"give me a heads up on the kentucky network \\ every 3 %hours"\end{tabular} & \begin{tabular}[c]{@{}c@{}}"show me the status for the last \\ five %prompts"\end{tabular} \\ \hline
% \begin{tabular}[c]{@{}c@{}}"I require a periodic summary of the \\ vancouver network every 8 %hours."\end{tabular} & "did the previous intent work" \\ \hline
% \end{tabular}
% \end{table}

\textit{2) Dynamic Route Integration}
After a function is defined, the semantic router parses the function details. These details are then passed to the associated static route definition, thus correlating the dynamic route with the previously defined static route. Lastly, a route layer is instantiated with the new static route details and the LLM of choice is passed as an argument in the route layer definition. When called, the new layer will facilitate the extraction of user intent alongside the extraction of details from the user’s prompt to create a key-value pairing that shows how the user wants their request fulfilled. The route layer used for dynamic route testing in this work relies on the OA encoder. Although encoder selection is important for accurate route selection, the key factor in dynamic route testing is not the choice of the encoder. \hl{Instead, the key factors for dynamic route testing are function definition and LLM choice. Properly defining a function and choosing an effective LLM to execute the function are critical to the extraction quality. Thus,} using the OA encoder for dynamic routes establishes a baseline to test how well detail extraction performs. 

\textbf{Experiments:}
\noindent\textit{1) Encoder Performance}
All five encoders were tested using the same set of RNR, IRR, and None route prompts. Each set of prompts \hl{was} generated with the \hl{aforementioned specifications and assumptions}. To test how well each encoder performs, before and after training, each prompt is given to the encoder's associated route layer with the expectation of triggering the prompt's associated route. For example, each RNR prompt that is sent to the route layer should trigger an RNR route. If the result is not the expected route, it is considered a failure of the encoder to properly extract the intent of the message. For each route layer, the expected result of each prompt is compared to the actual result from the encoder. This is then formed into a ratio of the encoder’s correct results compared to the total prompts in each set, effectively \hl{becoming} a binary classification problem where accuracy is used as the evaluation metric.

 A \textit{None Route} is triggered when the message does not meet the criteria of either the route layer’s RNR or IRR thresholds. These routes are specifically generated to have no intent or relevant request details. Because of this, this route can be thought of as a default choice for the route layer. Ensuring that the encoder effectively chooses this route when given an irrelevant prompt shows that the encoder understands the user’s intent. Notably, the None route further promotes the inherent guardrailing of the semantic router by ensuring that messages not intended for the 5GC do not trigger an intent route. 

\noindent\textit{2) Dynamic Route Performance}
To test the performance of the dynamic route, a series of prompts are sent to the route layer with specific details intended for extraction. Any prompt that is sent to the dynamic route layer that does not execute the expected static route is excluded from the results, as this experiment is not intended to test the performance of the encoder. Rather, the experiment detailed here aims to test how well the dynamic route layer interprets key information that the user provides for request execution. Similar to the encoder performance test, each prompt has an expected output with an expected format. This experiment focuses solely on the correctness of the route layers’ dynamic route output. 

\textit{Regular Notification Request}
A prompt destined for an RNR dynamic route is considered correct if its output properly extracts the frequency and subject of notification. The frequency is represented within the key-value results as its own key-value structure. This structure details values ranging from months to seconds. The result within this key-value structure is considered correct if it represents the amount of time listed by the user and, if needed, properly converts time values. Entries of 60 minutes, for example, should be converted to 1 hour. This ensures that a combined value such as 65 minutes is properly parsed as 1 hour and 5 minutes. Other conversions ensure that values of time are represented in the simplest form possible for easy processing in future implementations. The subject of notification is considered correct if the dynamic route properly lists the network element that the user inputs. 

\textit{Intent Report Request}
A prompt destined for an IRR dynamic route is successful if the key-value 
pair results in either the correct UUID-based intent identifier listed by the user or a list of negative numbers used as positional identifiers denoting \textit{n} previous intents. Any string of characters and numbers preceded by the “I-” prefix is considered a valid identifier for this dynamic route. A list of positional identifiers are the result of a phrase such as “4 last intents”, where the result in the dictionary object is a list of negative integers from -1 to -4. As previously stated, an example of IRR prompts and their output can be seen in Fig. \ref{fig:rnr_irr_examples}.

\section{Results and Analysis}

This section details the results of the aforementioned experiments. For each encoder, the accuracy and thresholds were compared before and after training. These analyses were split into three performance tests for RNR, IRR, and None routes.

\subsection{Static Route Performance}
\begin{figure}[!hbt]
\centerline{\includegraphics[width=0.75\columnwidth]{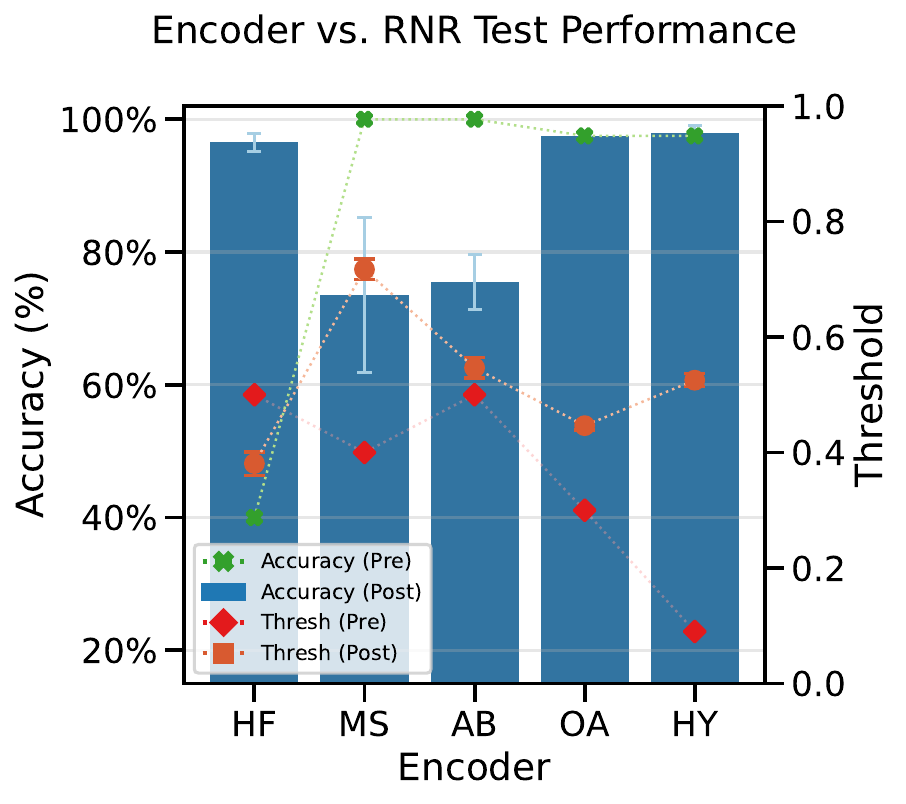}}
\caption{Post Training Encoder Comparative Analysis: RNR}
\label{fig:results_rnr}
\end{figure}

\textbf{RNR Performance:} Results of the RNR performance test can be seen in Fig. \ref{fig:results_rnr}.  In this figure, and all subsequent figures, the pre-/post-training accuracy and encoder threshold values across all experiments are shown. Post-training accuracies are reported as the mean and standard deviation across all trials to indicate average performance and stability. The HF encoder saw a significant improvement after training. However, the same cannot be said for MS and AB. Notably, the threshold change for HF, MS, and AB had a great effect on the accuracy of each. Initially, for both MS and AB, a low threshold allowed for a low barrier to route selection. After training, both encoders changed their thresholds to make route selection much more strict, thus creating a significant drop in accuracy. Out of all the encoders, two stand out in terms of performance: OA and the HY encoder. While the HY encoder performs slightly better than OA, the OA encoder remains the better choice for accuracy, given its more consistent performance, as evidenced by its lower standard deviation across experiments. The high accuracy and consistency of the OA encoder are observed throughout the results. 

\begin{figure}[!hbt]
\centerline{\includegraphics[width=0.75\columnwidth]{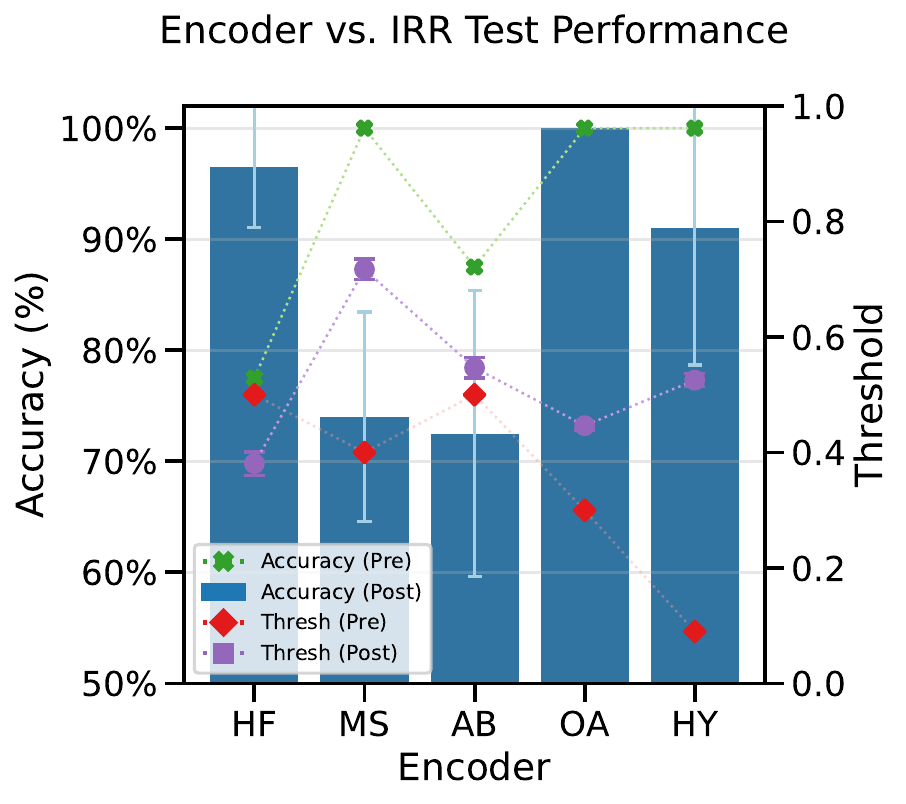}}
\caption{Post Training Encoder Comparative Analysis: IRR}
\label{fig:results_irr}
\end{figure}

\textbf{IRR Performance:} Results of the IRR performance test can be seen in Fig. \ref{fig:results_irr}. Similar to the RNR performance test, the HF encoder sees a great improvement in accuracy post-training. However, the stability in its accuracy across trials has greatly decreased. This is the case for all encoders except for the OA encoder. Again, a performance decrease can be seen in MS and AB along with a significant decrease in performance stability. While the HY encoder competed with the OA encoder’s performance before training, the HY encoder’s post training results show that OA is still the highest performing option. The HY encoder saw a large increase in threshold after training, making route selection harder, thus decreasing overall accuracy. Again, the performance of the OA shows high accuracy, low error, and consistency between route performance tests.

\begin{figure}[!hbt]
\centerline{\includegraphics[width=0.7\columnwidth]{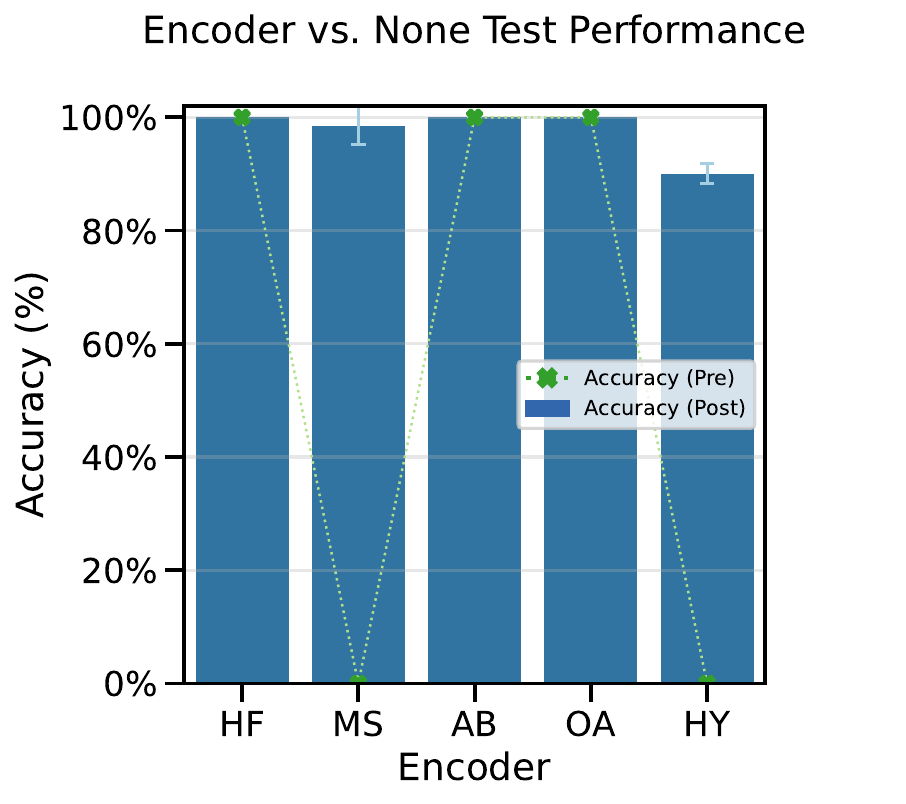}}
\caption{Post Training Encoder Comparative Analysis: None}
\label{fig:results_none}
\end{figure}

\textbf{None Performance:} In the case of the None route, thresholds are not needed as selection of the None route only takes place when the user prompt does not exceed the threshold of either the RNR or IRR for each encoder. The accuracy of the None route was tested by giving each encoder a series of messages that had no relevant intent. Results can be seen in Fig. \ref{fig:results_none}. Three encoders stand out in the results: HF, AB, and OA. All three encoders maintained high accuracy before and after training. At first, both MS and the HY encoder performed poorly. While post training results show an improvement for both, their instability and comparatively low accuracy become evident. The importance of None route accuracy cannot be understated in a live system. Without proper identification of irrelevant prompts, inaccurate values could be passed to fulfillment functions, causing unwanted or possibly damaging changes to the 5GC. 

\textbf{Overall Static Route Performance:} In each performance test, the OA encoder not only maintained high accuracy throughout training, but, most notably, it performed with high stability. Although the HY and the HF encoder exhibited comparable performance, OA’s robustness reinforces its high accuracy across all trials. In a live system, an encoder with high accuracy and low error would be vital for proper and safe fulfillment of intents in the 5GC. 

\subsection{Dynamic Route Performance}

Dynamic route execution performance was tested using a route layer with the OA encoder and the Mistral 7B LLM for function execution. Forty test samples for both RNR and IRR were passed to the route layer. The response for each \hl{sample prompt} was directly compared to the expected response. 

\textbf{RNR Performance:} Out of the forty \hl{samples} sent to the route layer, one message was misclassified and took the incorrect static route. In this case, this response was excluded as this experiment is specifically designed to test the performance of parameter extraction from user messages rather than route selection accuracy. Since static route selection performance has already been assessed, removing this sample ensures relevant data is represented in the results. Of the remaining samples, 38 of 39 were correctly extracted in their entirety, corresponding to an \textit{accuracy of 97.4\%}. When examining the single misclassified sample, it is worth noting that the extraction error which occurred was a conversion error. The LLM extracted a value of \texttt{‘minutes’:60}; however, the expected extraction value is \texttt{‘hours’:1, ‘minutes’:0}, due to explicit time-based conversion in the LLM prompt. It should be noted that this error was isolated as other sample values were converted successfully. Furthermore, post-processing checks would identify and correct conversion errors. While this error is not detrimental to system performance, it does motivate the need for LLM refinement, either through the use of a more robust base model, additional prompting, or fine tuning, something which will be the subject of future work.

\textbf{IRR Performance:} Of the forty user prompts sent to the route layer, only two messages resulted in an unexpected extraction output. This result corresponds to an \textit{accuracy of 95\%}. It should be noted that both of these errors were related to positional statements, and the extraction of intent IDs was 100\% successful across all relevant samples. Regarding the positional errors, the LLM was able to correctly identify the existence of a positional intent; however, the extracted intent interval was not entirely correct. For example, in one of the samples, the expected positional extraction, corresponding to the previous four intents, would have been \texttt{-1, -2, -3, -4}; however, the extracted value omitted the \texttt{-4} value. In the second sample, the expected extraction was \texttt{-1, -2}; however, values of \texttt{-2, -3} were extracted. These positional errors further support the need for additional enhancements. Future iterations will include post-processing, which will identify non-sequential positional parameters and correct them. 
% Alongside this, LLM refinement will ensure that all relevant positional parameters are included in the output. 

\subsection{Comparison to SOTA}
Past work has focused solely on the accuracy of static route selection without considering dynamic routes \cite{10901065}. Specifically, using the best-performing encoder, an accuracy of 94\% was achieved across a limited validation study. This work not only expands the validation study by considering the performance of route selection individually rather than collectively, but also increases the base performance of the semantic router. These considerations resulted in a \textit{collective accuracy exceeding 99\% with individual accuracies reaching 100\%}. This demonstrates a significant improvement in static route selection and successful refinement of semantic router performance. Furthermore, this work directly addresses a known limitation of past work by implementing dynamic routes and enabling reliable parameter extraction. These improvements work towards a comprehensive \textit{human-out-of-the-loop} network management framework.

% \begin{figure*}[!hbt]
%     \centering
%     \begin{subfigure}{0.32\textwidth}
%         \includegraphics[width=\textwidth]{Figures/RNR_Results_New_Utterances.pdf}
%         \caption{RNR Route}
%     \end{subfigure}
%     \begin{subfigure}{0.32\textwidth}
%         \includegraphics[width=\textwidth]{Figures/IRR_Results_New_Utterances.pdf}
%         \caption{IRR Route}
%     \end{subfigure}
%     \begin{subfigure}{0.32\textwidth}
%         \includegraphics[width=\textwidth]{Figures/None_Results_New_Utterances.pdf}
%         \caption{None Route}
%     \end{subfigure}
%     \caption{Post Training Encoder Comparative Analysis per Route}
%     \label{fig:route_resutls}
% \end{figure*}

% \begin{figure*}[!hbt]
%     \centering
%     \begin{subfigure}{0.45\textwidth}
%         \includegraphics[width=\textwidth]{Figures/RNR_Results_New_Utterances2.pdf}
%         \caption{RNR Route}
%     \end{subfigure}
%     \begin{subfigure}{0.45\textwidth}
%         \includegraphics[width=\textwidth]{Figures/IRR_Results_New_Utterances3.pdf}
%         \caption{IRR Route}
%     \end{subfigure}
%     \begin{subfigure}{0.45\textwidth}
%         \includegraphics[width=\textwidth]{Figures/None_Results_New_Utterances2.pdf}
%         \caption{None Route}
%     \end{subfigure}
%     \caption{Post Training Encoder Comparative Analysis per Route}
%     \label{fig:route_resutls}
% \end{figure*}

\section{Conclusion}
The work presented in this paper further realizes LLM-enhanced Intent-Based Networking in the 5G+ Core by introducing dynamic semantic routing to a previously proposed intent-extraction architecture. This advances intent fulfillment by extracting and formatting relevant prompt details. Intent extraction performance was tested using various encoders, including a hybrid encoder. The system effectively generates schema outputs and accurately extracts relevant information. 

Future work in this area will focus on further enhancing the proposed system model and implementing intent fulfillment. Since this work focused on schema output generation, a system model that ingests user inputs generated from the dynamic route process is the next logical step. A dedicated and virtualized infrastructure to fulfill intents is necessary, and integration with a live 5G+ network is required to fully realize the envisioned ZSM architecture.

\section*{Acknowledgments}
Research supported by the NVIDIA Academic Grant Program using DGX Spark.

% \begin{table}[htbp]
% \caption{Table Type Styles}
% \begin{center}
% \begin{tabular}{|c|c|c|c|}
% \hline
% \textbf{Table}&\multicolumn{3}{|c|}{\textbf{Table Column Head}} \\
% \cline{2-4} 
% \textbf{Head} & \textbf{\textit{Table column subhead}}& \textbf{\textit{Subhead}}& \textbf{\textit{Subhead}} \\
% \hline
% copy& More table copy$^{\mathrm{a}}$& &  \\
% \hline
% \multicolumn{4}{l}{$^{\mathrm{a}}$Sample of a Table footnote.}
% \end{tabular}
% \label{tab1}
% \end{center}
% \end{table}

%\begin{figure}[!hbt]
%\centerline{\includegraphics[width=0.9\columnwidth]{Figures/where_to_go.jpg}}
%\caption{Some days, other days, best days}
%\label{fig:mm}
%\end{figure}

\bibliographystyle{IEEEtran}
\bibliography{sample}

\end{document}